\documentclass[twocolumn,prb,aps,showpacs]{revtex4}
\usepackage[dvips]{graphicx}
\usepackage{graphicx}
\usepackage{dcolumn}
\usepackage{bm}

\begin{document}

\title{Tunable lateral displacement and spin beam splitter for ballistic
electrons in two-dimensional magnetic-electric nanostructures}

\author{Xi Chen $^{1}$\footnote{Email address: xchen@shu.edu.cn}}

\author{Chun-Fang Li $^{1,2}$\footnote{Email address: cfli@shu.edu.cn}}

\author{Yue Ban $^{1}$}

\affiliation{$^{1}$ Department of Physics, Shanghai University,
Shanghai 200444, People's Republic of China}

\affiliation{$^{2}$ State Key Laboratory of Transient Optics and
Photonics, Xi'an Institute of Optics and Precision Mechanics of CAS,
Xi'an 710119, People's Republic of China}

\date{\today}

\begin{abstract}

We investigate the lateral displacements for ballistic electron
beams in a two-dimensional electron gas modulated by metallic
ferromagnetic (FM) stripes with parallel (P) and anti-parallel (AP)
magnetization configurations. It is shown that the displacements are
negative as well as positive, which can be controlled by adjusting
the electric potential induced by the applied voltage and the
magnetic field strength of FM stripes. Based on these novel
phenomena, we propose an efficient way to realize a spin beam
splitter, which can completely separate spin-up and spin-down
electron beams in the AP configuration by their corresponding
spatial positions.



\end{abstract}

\maketitle

It is well established that ballistic electrons in the
two-dimensional electron gas (2DEG) are reflected, focused,
diffracted, and interfered in a manner similar to the
electromagnetic waves in dielectrics
\cite{Gaylord-Brennan-Gaylord-G-H,Wilson-G-G}, which results from
the quantum-mechanical wave nature of electrons, thus have given
rise to a field of research which is best described as ballistic
electron optics in 2DEG systems. In the past two decades, many
electronic analogues of optical devices \cite{Dragoman-D,Datta} have
been studied, including electron gratings, electron waveguide,
electron interferometer, and electron wave beam splitter .

Recently, electron spin beam splitter and spin filter in different
2DEG systems have been more attractive
\cite{Kiselev,Shelykh,Foldi,Khodas,Ramaglia,Zhang} for the nascent
field called ``Spintronics" \cite{Zutic}, which is a
multidisciplinary field whose central subject is the active control
and manipulation of spin degree of freedom in solid systems. For
example, Khodas \textit{et al.} \cite{Khodas} have proposed the
basic schemes for filtration and control of the electron spin by
electron spin optics in the 2DEG structures with Rashba and
Dresselhaus spin-orbit coupling. Moreover, spin beam splitters or
spatially separating spin filters have also been investigated by
other optics-like phenomena,respectively, such as spin double
refraction \cite{Ramaglia} and negative refraction \cite{Zhang}.
However, considering some disadvantages of the 2DEG structures with
spin-orbit coupling, Frustaglia \textit{et al.} \cite{Frustaglia}
have studied spin filters by pure quantum interference effect.
Dragoman \cite{Dragoman} has lately presented an alternative spin
beam splitter in 2DEG systems, in terms of magnetic depopulation of
sub-bands in magnetic fields.

In this Brief Report, we will investigate the lateral displacements
for ballistic electron beams in a 2DEG modulated by metallic
ferromagnetic (FM) stripes with parallel (P) and anti-parallel (AP)
magnetization configurations. The displacements can be negative and
positive, which can be controlled by adjusting the total electric
potential and the magnetic field strength of FM stripes. More
interestingly, we propose a spin beam splitter as a potential
application of an intriguing phenomenon in which, large and opposite
lateral displacements may occur simultaneously for spin-down and
spin-up electron beam in the AP configurations. As a matter of fact,
The displacements are closely related to the Goos-H\"{a}nchen (GH)
effect in optics \cite{Goos}, and are the electronic analogous to
the simultaneously large and opposite generalized GH shifts for TE
and TM light beams in an asymmetric double-prism configuration
\cite{Li}. Thus, the tunable negative or positive displacement
introduced here provides a completely different mechanism of
spin-polarized electron splitting, also at the nanoscale level but
with more design flexibility.

\begin{figure}[]
\scalebox{0.70}[0.70]{\includegraphics{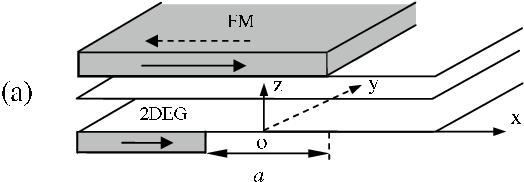}}\\
\scalebox{0.62}[0.62]{\includegraphics{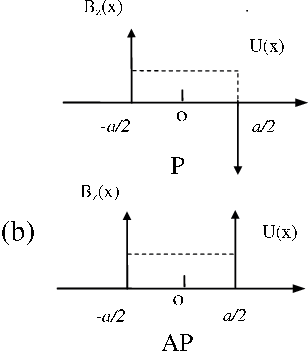}}
\scalebox{0.62}[0.62]{\includegraphics{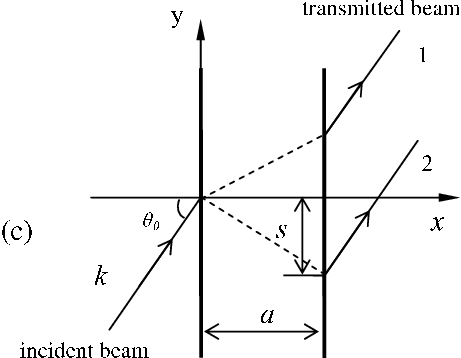}}
 \caption{(a) Schematic illustration of the magnetic-electric
nanostructure with two metallic FM stripes deposited on top and
bottom of the semiconductor heterostructure; (b) The
magnetic-electric barrier models exploited here corresponds to the P
and AP magnetization configurations of two FM stripes, respectively.
(c) The positive (1) and negative (2) lateral displacements of
ballistic electron beams in this structure.} \label{fig.1}
\end{figure}

The system under consideration is a magnetically modulated 2DEG
formed usually in a modulation-doped semiconductor heterostructure
\cite{Kubrak-Vancura}, which can be experimentally realized by
depositing two metallic ferromagnetic (FM) stripes on top and bottom
of the semiconductor heterostructure, as schematically depicted in
Fig. \ref{fig.1}(a). For the small distances between 2DEG and FM
stripes, the magnetic field provided by two FM stripes is
approximated \cite{Papp-Peeters} as $B_z(x)=[B_1 \delta(x+a/2)- \chi
B_2 \delta(x-a/2)]$, and the total electric potential induced by the
negative voltage applied directly to the 2DEG
$U(x)=U\Theta(a/2-|x|)$ are homogenous in the $y$ direction and vary
only along the $x$ axis \cite{Lu}, which is shown in Fig.
\ref{fig.1}(b), where $\chi$ represents the parallel (P) and
anti-parallel (AP) magnetization configuration of two FM stripes
($\chi=+ 1$ for P and $\chi =-1$ for AP). The Hamiltonian describing
such a system in the $(x, y)$ plane, within the single particle
effective mass approximation, is
\begin{equation}
H=\frac{p_x^2}{2m^*}+\frac{[p_y+e{\bf
A}_y(x)]^2}{2m^*}+U(x)+\frac{eg^*}{2m_0}\frac{\sigma \hbar}{2}
B_{z}(x),
\end{equation}
where $m^*$ is the electron effective  mass and $m_0$ is the free
electron mass, $(p_x, p_y)$ are the components of the electron
momentum, $g^*$ is the effective Land\'{e} factor of
electron, $\sigma=+1/-1$ for spin up/down electrons, and ${\bf
A}_y(x)$ is the y-component of the vector potential given, in Landau
gauge, by $\vec{{\bf A}}=[0, {\bf A}_y(x), 0]$. We express all the
relevant quantities in dimensionless form: $(1)$ the magnetic field
$B_{z}(x)\rightarrow B_0 B_{z}(x)$, $(2)$ the vector potential ${\bf
A}(x)\rightarrow B_0 l_B {\bf A}(x)$, $(3)$ the coordinates ${\bf r}
\rightarrow l_B {\bf r} $, $(4)$ the energy $E \rightarrow
E\hbar\omega_c$, where $\omega_c=eB_0/m^*$ is the cyclotron
frequency and $l_B=\sqrt{\hbar /e B_0}$ is the magnetic length with
$B_0$ as some typical magnetic field. For $GaAs$ and an estimated
$B_0=0.1 T$ we have $l_B=813 nm$, $\hbar\omega_c=0.17 meV$,
$g^*=0.44$, and $m^*=0.067m_0$.

A two-dimensional electron beam of incidence energy $E$ comes from
the left with an incidence angle $\theta_0$ in $(x,y)$ plane, as is
depicted in Fig. \ref{fig.1}(c). Let $
\Psi_{i}(\vec{x})=A(k_y)\exp\{i[ k^{l}_{x} (x+a/2)+k_y y]\}$ be the
plane wave component of the incident beam, where $k^{l}=\sqrt{2E}$,
$k_y=k^{l}\sin\theta$, $k^{l}_{x}=k^{l} \cos\theta$, $\theta$ stands
for the incidence angle of the contributed plane wave, and $A(k_y)$
is the angular-spectrum distribution. Because the system is
translational invariant along the $y$ direction, the solution of the
stationary Schr\"{o}dinger equation $H\Psi(x,y)=E\Psi(x,y)$ can be
written as $\Psi(x,y)=\psi( x)\exp{(i k_y y)}$. The wave function
$\psi(x)$ satisfies the following one-dimensional Schr\"{o}dinger
equation:
\begin{equation}
\label{Schrodinger equation}
\left\{\frac{d^2}{dx^2}-2\left[E-U_{eff}(x,k_y,\sigma)\right]\right\}\psi(x)=0,
\end{equation}
where the effective potential in the barrier region
$U_{eff}(x,k_y,\sigma)=U+[k_y+A_y(x)]^2/2+m^*g^*\sigma B_z(x)/4m_0$
depends not only on the wave vector $k_y$, but also on the
interaction between the electron spin and the nonhomogeneous
magnetic field. According to Eq. (\ref{Schrodinger equation}) and
boundary conditions, that is, the continuity of the wave functions
and their derivations at the boundaries, the wave function of the
corresponding plane wave of transmitted beam is found to be $
\label{transmitted wave} \Psi_t (\vec{x})= t(k_y) A(k_y) \exp
\{i[k^{r}_{x} (x-a/2) +k_y y]\}, $ and the amplitude transmission
coefficient $t(k_y)=e^{i \phi}/g$ is determined by the following
complex number, $ ge^{i \phi}= 2 k^{r}_{x} k'_{x}/(M + i N),$ where
\begin{widetext}
\begin{eqnarray}
\label{M} \label{M} M=k'_x (k^{l}_x +k^{r}_{x}) + \frac{m^{*}g\sigma
}{2 m_0}(k^{l}_{x} B_2 - \chi k^{r}_{x}B_1)\tan k'_x a,
\end{eqnarray}
\begin{eqnarray}
\label{N} \label{N} N=-\frac{m^{*}g\sigma}{2 m_0} (B_1-\chi B_2)
k'_x +\left[k^{l}_{x} k^{r}_{x}+k'^2+\left(\frac{m^{*}g\sigma }{2
m_0}\right)^2(\chi B_1 B_2)\right]\tan k'_x a,
\end{eqnarray}
\end{widetext} so that the total phase shift of the transmitted
beam at $x=a/2$ with respect to the incident one at $x=-a/2$ is
$$\tan \phi \equiv \frac{N}{M}, $$ where $k'_x=[2E-(k_y+B_1)^2]^{1/2}$ and
$k^{r}_{x}=[2E-(k_y+B_1-\chi B_2)^2]^{1/2}$. As indicated in Fig.
\ref{fig.1} (c), the lateral displacement of the transmitted beam is
defined, according to the stationary phase approximation
\cite{Wilson-G-G,Bohm,Chen}, as
\begin{equation}
\label{definition} s=-d\phi/dk_{y0},
\end{equation}
where the subscript $0$ in this paper denotes the values taken at
$k_y=k_{y0}$, namely, $\theta=\theta_0$. It is clearly seen from
Eqs. (\ref{M}) and (\ref{N}) that the lateral displacement presented
here is dependent of the electron spin, except only when $B_1=B_2$
in the P case of $\chi=1$. For the simplicity, we will let
$B_1=B_2=B$ in the following discussions on the lateral displacement
in the P and AP configurations, respectively.

Firstly, the lateral displacement (\ref{definition}) in the P case
is reduced to,
\begin{widetext}
\begin{eqnarray}
\label{lateral displacement} s &=& \frac{s_{g}}{2 g^2_0}
\left\{\left( \frac{k_{x0}}{k'_{x0}}+
\frac{k'_{x0}}{k_{x0}}+\frac{k_0^2}{k_{x0}k'_{x0}}\right) -
\left[\left(1-\frac{k'^2_{x0}}{k_{x0}^2}\right)\left(\frac{k_{x0}}{k'_{x0}}-
\frac{k'_{x0}}{k_{x0}}\frac{k_{y0}}{k'_{y0}}\right)+\frac{k_0^2}{k_{x0}^2}\left(\frac{k_{x0}}{k'_{x0}}+
\frac{k'_{x0}}{k_{x0}}\frac{k_{y0}}{k'_{y0}}\right)\right]
\frac{\sin 2k'_{x0} a }{2k'_{x0} a}\right\},
\end{eqnarray}
\end{widetext}
where $k_{x0}=(2E-k^2_{y0})^{1/2}$, $k_0=m^{*}g\sigma B/2 m_0$,
$k'_{y0}=k_{y0}+B$, $s_{g}= a \tan \theta'_0$, $\tan
\theta'_0=k'_{y0}/k'_{x0}$, and the transmission probability
$1/g^2_0=T_0$ is closely related to the measurable ballistic
conductance $G$, according to the well-known Landauer-B\"{u}ttiker
formula \cite{Buttiker}.

Eq. (\ref{lateral displacement}) indicates that the modulation of
the lateral displacement $s$ relies on the following properties: (1)
When transmission resonances occur, that is to say, when $k'_{x0} a=
m \pi (m=1,2,3,...)$ is satisfied so that $T_0=1$. At resonances,
the lateral displacement is also maximal,
$$
s_{max} \equiv s|_{k'_{x0} a= m \pi}=
\frac{s_{g}}{2}\left(\frac{k_{x0}}{k'_{x0}}+
\frac{k'_{x0}}{k_{x0}}+\frac{k_0^2}{k_{x0}k'_{x0}}\right).
$$
The resonance condition depends on the electric potential $U$ and
magnetic field strength $B$. (2)  The lateral displacement can be
negative under the necessary condition, which can be expressed as a
restriction to the incidence angle $\theta_0$ as follows:
\begin{equation}
\label{restriction to incident angle} \cos \theta_0 <
\left[\frac{-U-(B^2-k_0^2)/2}{2E}\right]^{1/2} \equiv \cos \theta_t,
\end{equation} because of the factor $\sin 2k'_{x0} a
/2k'_{x0} a \leq 1$. This shows that if the incidence angle
satisfies the condition (\ref{restriction to incident angle}), that
is to say, if $\theta_0$ is larger than the threshold angle
$\theta_t$, one can always find a width $a$ where the displacement
is negative. Indeed, Eq. (\ref{restriction to incident angle}) isn't
satisfied by any incidence angle if $U > 0$, since $B^2> k_0^2$.
That means that the displacement is always positive for $U>0$, while
it can be negative for $U<0$. The lateral displacement can thus be
easily tuned from positive to negative by adjusting the electric
potential $U$, with changing the corresponding negative and positive
voltages applied directly to 2DEG.

Further investigations show that Eq. (\ref{lateral displacement}) is
still valid in the evanescent case of $2(E-U)<(k_y+B)^2$, only if
$k'_x$ is replaced by $i \kappa$, where
$\kappa=[(k_y+B)^2-2(E-U)]^{1/2}$. The displacement in this case
saturates to a positive constant in the opaque limit $ a \gg
1/\kappa_0$ with the transmission probability decaying exponentially
in the same ways as that in the semiconductor barrier structure
\cite{Chen}.

\begin{figure}[]
\scalebox{0.32}[0.3]{
 \includegraphics{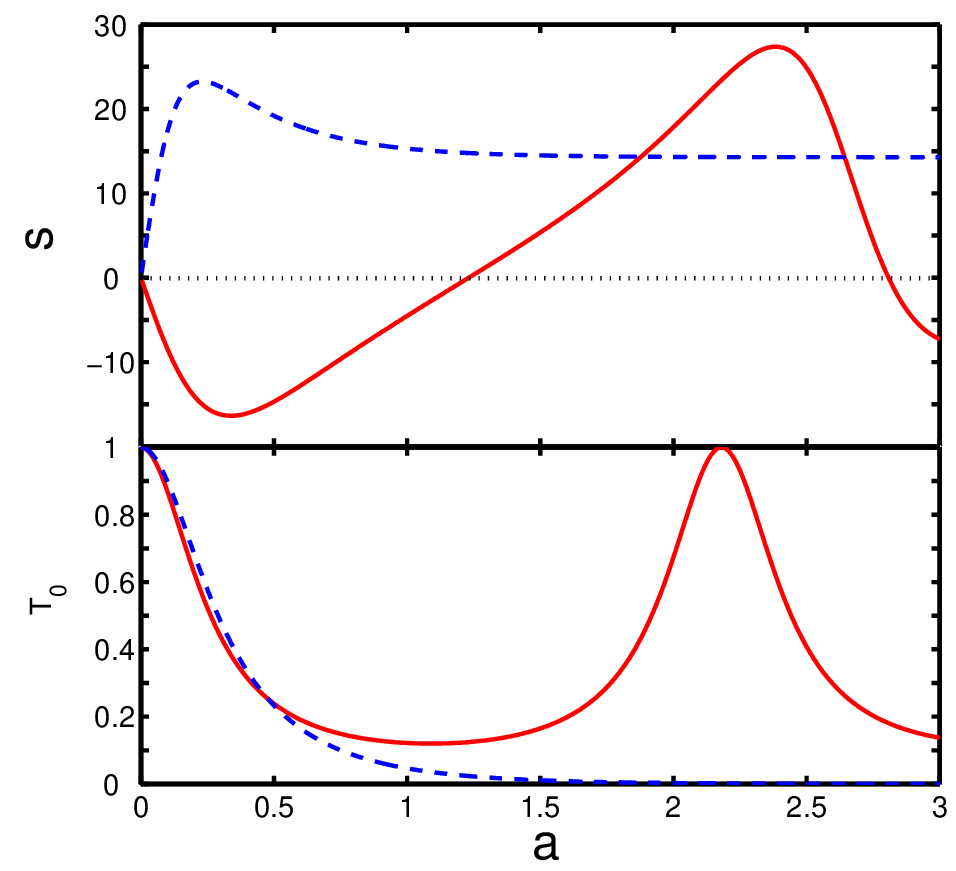}}
\caption{(Color online) Dependence of the displacements and the
corresponding transmission probabilities on the width $a$ for $U=1$
(solid curve) and $U=-1$ (dashed curve), respectively, where $E=5$,
$B=0.1$, $\theta_0=85^{\circ}$.} \label{fig.2}
\end{figure}
\begin{figure}[]
\scalebox{0.28}[0.28]{
  \includegraphics{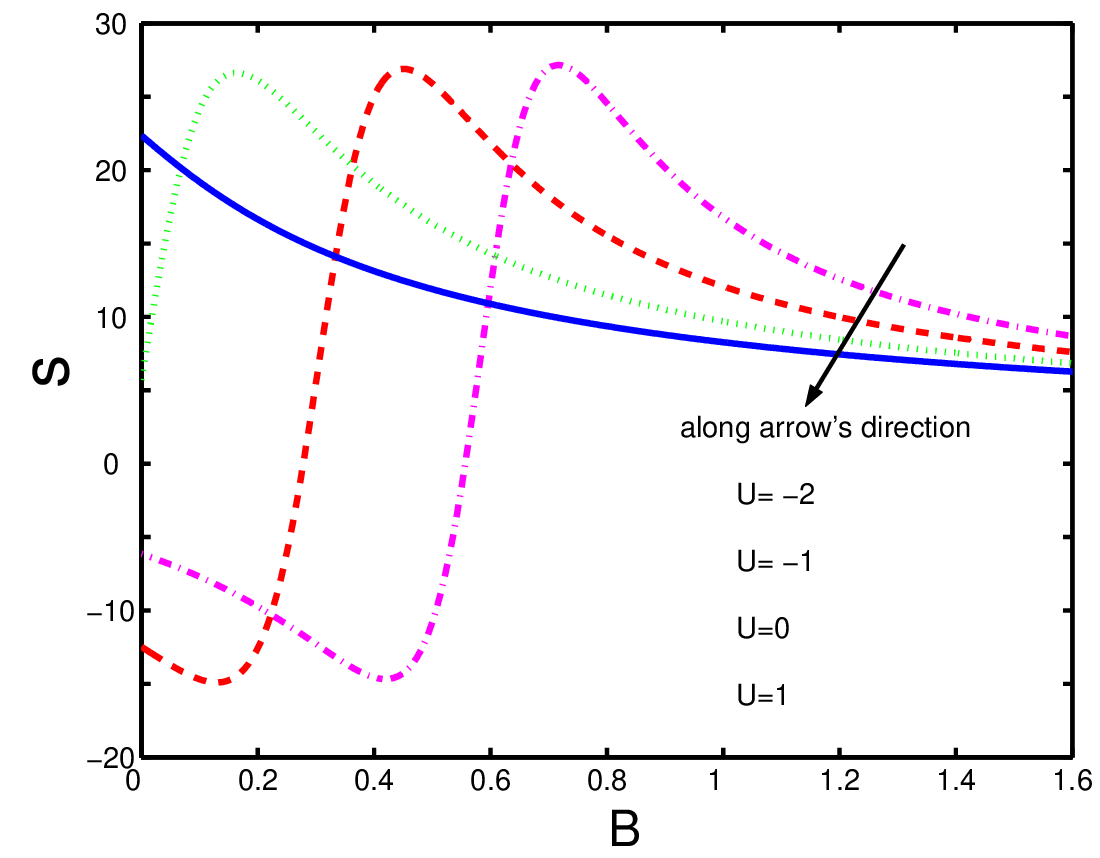}}
  \caption{(Color online) Modulation of the displacements
  by the magnetic field strength $B$ at different fixed electric
  potential $U=-2$ (dotted-dashed curve), $-1$ (dashed curve), $0$
  (dotted curve) , and $1$ (solid curve), respectively, where $a=0.5$
and other parameters are the same as in Fig. \ref{fig.2}.}
\label{fig.3}
\end{figure}

In Fig. \ref{fig.2} is shown that the typical dependence of the
lateral displacement (\ref{lateral displacement}) and the
corresponding transmission probability on the width $a$ for $U=1$
(dash curve) and $U=-1$ (solid curve), respectively, where $E=5$,
$B=0.1$, $\theta_0=85^{\circ}$($\theta_t=71.5^{\circ}$), namely,
$k_{y0}=3.2$. The displacement and transmission probability are
identical for spin-up and spin-down electrons in this case. It is
also shown that the displacement depends periodically on the width
$a$ in the propagating case, while it saturates to a constant for an
opaque barrier in the evanescent case. Calculations under those
conditions show that the displacements for $U=-1$ and $U=1$ are
approximately equal to $-12.48$ and $22.36$ for $a=0.5$,
respectively. This implies that the lateral displacements can be
opposite for the different sign of $U$ with the approximately equal
amplitude of the corresponding transmission probabilities.

Fig. \ref{fig.3} presents the modulation of the lateral
displacements by the magnetic field strength $B$ at the fixed
electric potential $U$. From these four curves, one can observe the
absorbing feature that the displacements can be changed from
negative to positive with the increasing $U$. Another observation
from Fig. \ref{fig.3} is that, with the increasing $B$, the curve of
lateral displacement shifts leftwards. Moreover, when $U$ and $B$
are large enough, the lateral displacements always tend to positive
values in the evanescent case. All those features result from the
dependence of the effective potential $U_{eff}$ on  $U$ and $B$.
\begin{figure}[]
\scalebox{0.22}[0.25]{\includegraphics{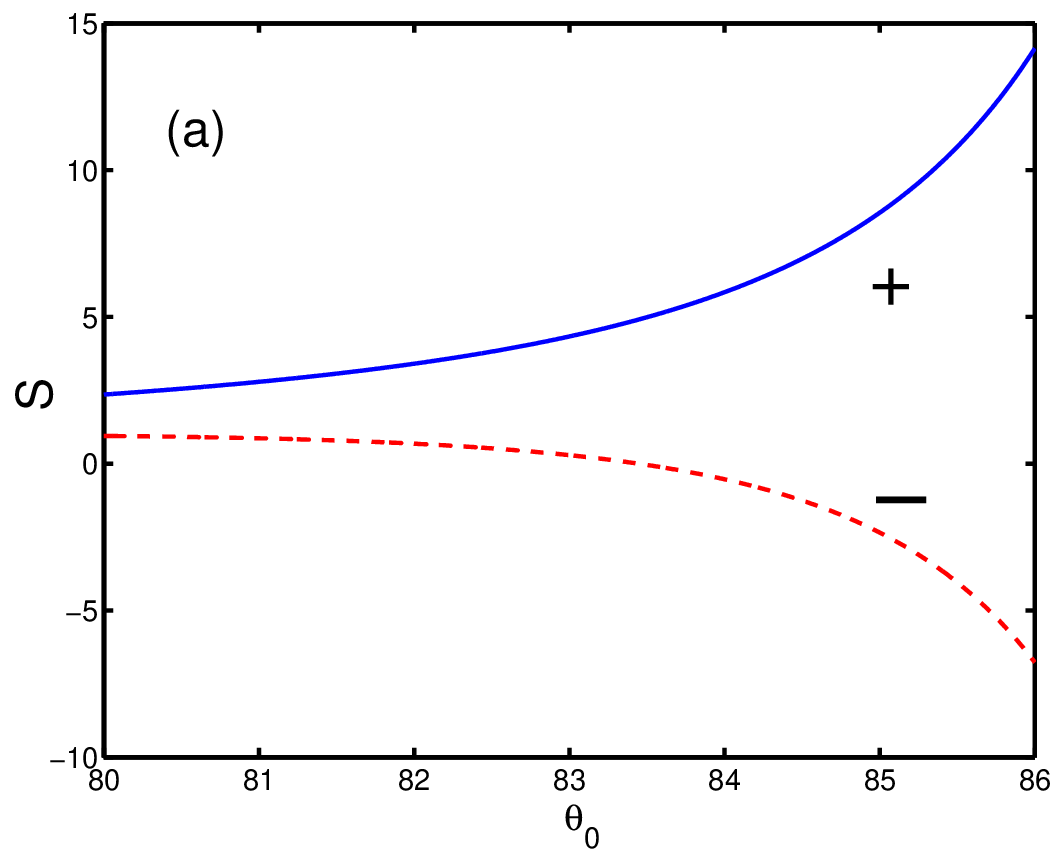}}
\scalebox{0.22}[0.25]{\includegraphics{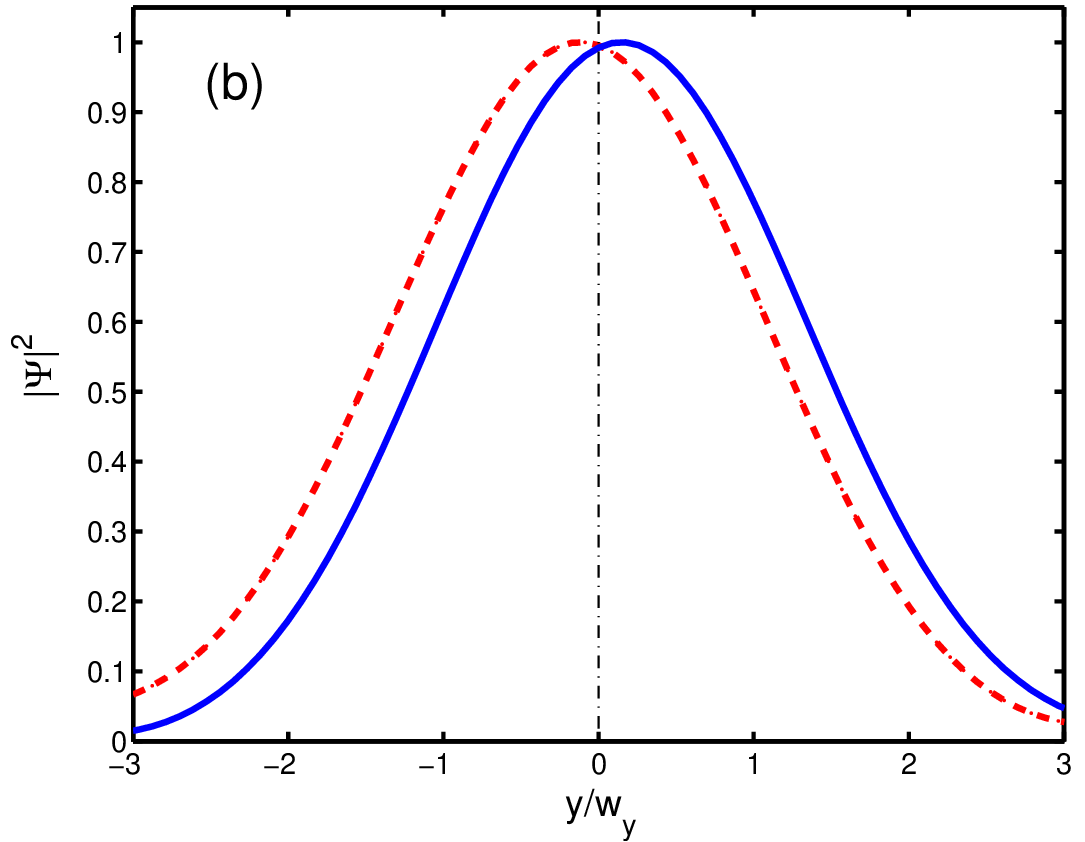}} \caption{(Color
online) (a) Simultaneously large and opposite lateral displacements
for spin-up (solid curve) and spin-down (dashed curve) electron
beams, where $E=2$, $a=0.5$, $U=-1$ and the other parameters are the
same as in Fig. \ref{fig.2}. (b) Comparison of the corresponding
normalized shapes of the transmitted beams.} \label{fig.4}
\end{figure}
This leads to the modulations of the displacement by the total
electric potential induced by the voltage applied directly to 2DEG
and magnetic field strength produced by FM stripes.

Now, we are ready to investigate the a spin-polarized ballistic
electron beam splitter based on the properties of the lateral
displacements discussed above. In the AP configuration, the lateral
displacement  depends on the electron spin due to its asymmetry
\cite{Zhai-Xu}, so that the spin-polarized electron beam can be
separated spatially. More interestingly, Fig. \ref{fig.4} (a) gives
an example that the displacement can be opposite for spin-up and
spin-down electron beams at a large incidence angle, where $E=2$,
$a=0.5$, $U=-1$ and the other parameters are the same as in Fig.
\ref{fig.2}. Fig. \ref{fig.4}(b) further demonstrates the validity
of the above stationary-phase analysis by numerical simulations of
Gaussian-shaped incident beam, that is to say, the transmitted beam
retains well the shape of the incident beam with positive and
negative displacements, within the restriction $a \ll \pi w/(2 \cos
\theta_0 \tan\theta_0')$ \cite{Chen}, where $w=10 \lambda_e$. As
mentioned above, the positive displacement under these physical
parameters corresponds to the evanescent case, thus the
displacements presented here are not large enough, which leads to
the quite small spin beam spilt. However, further investigations
show that the opposite displacements for the electron beams
reflected from this system can be greatly enhanced by the
transmission resonance in the same way as the generalized GH shifts
for light beams in a double prism configuration \cite{Li}. In a
word, the control of the simultaneously large and opposite lateral
displacements allows this system to realize the spin beam splitter,
which can completely separate spin-up and spin-down electron beams
in the AP configuration by their different spatial positions.

In conclusion, we have investigated theoretically and numerically
the tunable lateral displacement and spin beam splitter for
ballistic electrons in magnetic-electric nanostructure. The
displacement presented here has the feature of GH shift, which does
result from the reshaping process of the transmitted beam, because
of the destructive and constructive interferences between each plane
wave components undergoing the different phase shift due to the
multiple reflections in this system \cite{Chen}. Recent
investigations show that the lateral GH displacement and transverse
displacement called Imbert-Fedorov (IF) effect relate directly to
spin Hall effect \cite{Onoda}. Thus, the spin-dependent
displacements and related phenomena in various quantum systems,
especially in presence of spin-orbit coupling, remain as further
problems. We hope that these interesting phenomena may stimulate
experiments to realize novel electronic devices, such as spin filter
and spin beam splitter.

This work was supported in part by the Shanghai Educational
Development Foundation (2007CG52), National Natural Science
Foundation of China (60377025), and the Shanghai Leading Academic
Discipline Program (T0104).

\end{document}